\documentclass[journal]{new-aiaa}
\usepackage[utf8]{inputenc}

\usepackage{graphicx,amssymb,stmaryrd,array,multirow,color,epstopdf}
\usepackage[table]{xcolor}
\usepackage{amsmath}
\usepackage[version=4]{mhchem}
\usepackage{siunitx}
\usepackage{longtable,tabularx}
\newcolumntype{M}[1]{>{\centering\arraybackslash}m{#1}}

\definecolor{gray}{rgb}{0.5, 0.5, 0.5}
\definecolor{red}{rgb}{0.8500, 0.1250, 0.0480}
\definecolor{blue}{rgb}{0, 0.4470, 0.7410}
\definecolor{green}{rgb}{0.4660, 0.6740, 0.1880}

\title{Resolvent Analysis of Compressible Laminar and Turbulent Cavity Flows}

\author{Yiyang Sun\footnote{Postdoctoral Research Associate, Department of Aerospace Engineering and Mechanics (UMN); Department of Mechanical Engineering (FSU), AIAA member, sun00249@umn.edu.}}
\affil{University of Minnesota, Minneapolis, MN 55455\\
Florida State University, Tallahassee, FL 32310}
\author{Qiong Liu\footnote{Postdoctoral Research Associate, Department of Mechanical and Aerospace Engineering (UCLA); Department of Mechanical Engineering (FSU), AIAA member, qliu3@g.ucla.edu.}}
\affil{University of California, Los Angeles, CA 90095\\
Florida State University, Tallahassee, FL 32310}
\author{Louis N.~Cattafesta III\footnote{Eminent Scholar and Professor, Department of Mechanical Engineering, AIAA Associate Fellow, lcattafesta@fsu.edu.}}
\affil{Florida State University, Tallahassee, FL 32310}
\author{Lawrence S.~Ukeiley\footnote{Associate Professor, Department of Mechanical and Aerospace Engineering, AIAA Associate Fellow, ukeiley@ufl.edu.}}
\affil{University of Florida, Gainesville, FL 32611}
\author{Kunihiko Taira\footnote{Associate Professor, Department of Mechanical and Aerospace Engineering (UCLA), Department of Mechanical Engineering (FSU), AIAA Associate Fellow, ktaira@seas.ucla.edu.}}
\affil{University of California, Los Angeles, CA 90095}
\affil{Florida State University, Tallahassee, FL 32310}

\usepackage{mdwlist}

\makecompactlist{itemize}{stditemize}

\usepackage{wrapfig}

\long\def\symbolfootnote[#1]#2{\begingroup%
\def\thefootnote{\fnsymbol{footnote}}\footnote[#1]{#2}\endgroup}

\begin{document}

\maketitle


\begin{abstract}

The present work demonstrates the use of resolvent analysis to obtain physical insights for open-cavity flows. Resolvent analysis identifies the flow response to harmonic forcing, given a steady base state, in terms of the response and forcing modes and the amplification gain. The response and forcing modes reveal the spatial structures associated with this amplification process.  In this study, we perform resolvent analysis on both laminar and turbulent flows over a rectangular cavity with length-to-depth ratio of $L/D=6$ at a free stream Mach number of $M_\infty=0.6$ in a spanwise periodic setting.  Based on the dominant instability of the base state, a discount parameter is introduced to resolvent analysis to examine the harmonic characteristics over a finite-time window.  We first uncover the underlying flow physics and interpret findings from laminar flow at $Re_D = 502$. These findings from laminar flow are extended to a more practical cavity flow example at a much higher Reynolds number of $Re_D = 10^4$.  The features of response and forcing modes from the laminar and turbulent cavity flows are similar to the spatial structures from the laminar analysis. We further find that the large amplification of energy in flow response is associated with high frequency for turbulent flow, while the flow is more responsive to low frequency excitation in the laminar case. These findings from resolvent analysis provide valuable insights for flow control studies with regard to parameter selection and placement of actuators and sensors. 

\end{abstract}

\section*{Nomenclature}

{\renewcommand\arraystretch{1.0}
\noindent\begin{longtable*}{@{}l @{\quad=\quad} l@{}}
$C_p$  & Pressure coefficient\\
$f_n$  & Rossiter frequency ($n$th)\\
$L,~\Lambda,~D$ & Cavity length, width, and depth\\
$M$  & Mach number \\
$p$  & Pressure \\
$e$  & Energy \\
$\mathcal F$  & Set of forcing modes\\
$\mathcal Q$  & Set of response modes\\
$W$  & Weight matrix\\
$Q$  & $Q$-criterion\\
$Re_D$  & Depth-based Reynolds number\\
$Re_\theta$  & Momentum-thickness-based Reynolds number\\
$St_L$  & Length-based Strouhal number \\
$t$  & Time \\
$u,~v,~w$  &  Streamwise, transverse and spanwise velocity\\
$x,~y,~z$  & Streamwise, transverse, and spanwise directions\\

$\rho$  & Density\\
$\tilde \alpha$  & Phase delay \\
$\beta$ & Spanwise wavenumber \\
$\gamma$  & Specific heat ratio \\
$\delta_0$  & Boundary layer thickness at cavity leading edge\\
$\alpha$ & Time scale parameter in discounted resolvent operator\\
$\theta_0$ & Momentum thickness at cavity leading edge\\
$\kappa$ & Averaged convection speed of disturbance \\
$\sigma$ & Singular value \\
$\omega_r$ & Radian frequency\\
$\omega_i$ & Growth/decay rate\\

\end{longtable*}}

\section{Introduction}
\label{sec:intro}

Modal analysis techniques are valuable tools to extract dominant features from a wide range of flows.  There are approaches that can identify spatial modes associated with flow unsteadiness, dynamics, instabilities, and harmonic responses \cite{Holmes96, Schmid:JFM10, Theofilis:ARFM11, Trefethen578, Taira_etal:AIAAJ17, Taira_etal:AIAAJ19}.  Resolvent analysis is one of these modal analysis techniques that can examine the harmonic input-output characteristics about a given base state \cite{Trefethen578, Jovanovic:JFM05}.  Closely related to the resolvent analysis is the global stability analysis, which identifies instabilities in the flow.  While the global stability analysis reveals how perturbations in the flow behave through the setup of an initial value problem, the resolvent analysis uncovers the flow response as a particular solution for a sustained harmonic forcing input.  The insights gained from resolvent analysis are powerful and have supported studies focusing on transitions, transient growth, and flow control \cite{Bagheri:AMR09, mckeon2010critical, Luhar:JFM14, Yeh:JFM19}.  

In fluid mechanics, resolvent analysis examines how harmonic forcing inputs can be amplified through the linearized Navier--Stokes operator with respect to the given base state.  Strictly speaking, the base state should be a steady state solution to the Navier--Stokes equation \cite{Trefethen578, Jovanovic:JFM05, Jovanovic:04}.  However, the nonlinear terms in the Navier--Stokes equations can be considered as part of the harmonic forcing input to the linearized system enabling the resolvent analysis to examine base states which are not the exact solution to the Navier--Stokes equation.  This extension requires the base flow to be statistically stationary such that the fluctuations from the nonlinear physics are comprised of harmonic inputs \cite{Farrell:PRL94, mckeon2010critical}.  These points enable resolvent analysis to examine time-average statistically stationary turbulent flows, making it very useful beyond traditional operator-based modal analysis techniques that focused on laminar flows.

In this work, we consider the application of resolvent analysis to compressible cavity flows. Rectangular cavities are ubiquitous in various engineering settings from small to large scales, including gaps between plates, automobile sunroofs, landing gear wells, and aircraft weapon bays. Flows over such cavities are known to exhibit high levels of fluctuations. In cavity flow, a shear layer emanates from the cavity leading edge, rolls up into large vortices from the Kelvin--Helmholtz instability, and impinges on the cavity aft-wall, producing intense pressure and velocity fluctuations as well as noise emission. These fluctuations can be large and cause structural fatigue and sound pollution. In an effort to address these issues, cavity flows have been studied for decades to understand the flow characteristics influenced by various flow conditions \cite{Lawson:PAS11}. For cavity flows, resolvent analysis has been performed on laminar flow conditions for the investigation of frequency selection mechanism \cite{Qadri:PRF17} and to design actuation strategies to suppress flow oscillations \cite{Liu:AIAA18}. The present paper extends resolvent analysis to turbulent flows and lays out the procedure as part of the special issue on modal analysis. 

The instabilities of cavity flows have been investigated in several studies over the past decades focusing on the effects of cavity aspect ratio, free stream Mach number, and sidewalls \cite{Bres:JFM08,Meseguer:JFM14,Citro:2015kj,Liu:JFM16,Sun:JFM17}. The findings provide rich information in terms of spatial distribution, growth/decay rate and temporal frequencies of disturbances that develop under the given base flows.  The fruitful results from stability analysis have also served as guidelines in many steady control efforts aimed at suppressing cavity flow oscillations \cite{Zhang:AIAAJ19,Sun:AIAAJ19}.  The resolvent analysis discussed in the present work is aimed to provide insights for unsteady control designs in turbulent flow to potentially improve the control performance. 

In the present work, resolvent analysis is performed on both laminar and turbulent cavity flows at $Re_D=502$ and $10^4$, respectively, to uncover the underlying physics. Stability analysis has also been conducted to examine the instabilities of the statistically stationary mean flow, which also provides guidelines for choosing an appropriate time-window in the discounted resolvent analysis. Moreover, we compare the findings from laminar and turbulent flows, which are valuable in understanding the underlying physics from a fundamental study at low Reynolds number to practical engineering applications at high Reynolds number. The paper is organized as follows. The methodologies of resolvent analysis and stability analysis are provided in section \ref{sec:Appro}, along with the computational setup of the open-cavity flow. In section \ref{sec:result}, we discuss the characteristics of the base flows obtained from direct numerical simulation and large eddy simulation, and present the two-dimensional (2D) and three-dimensional (3D) eigenmodes and resolvent modes from stability analysis and resolvent analysis, respectively.  At last, conclusions are provided in section \ref{sec:sum}.

\section{Approach}
\label{sec:Appro}

\subsection{Resolvent analysis}

\begin{figure}[hbpt]
\centering
\includegraphics[width=0.99\textwidth]{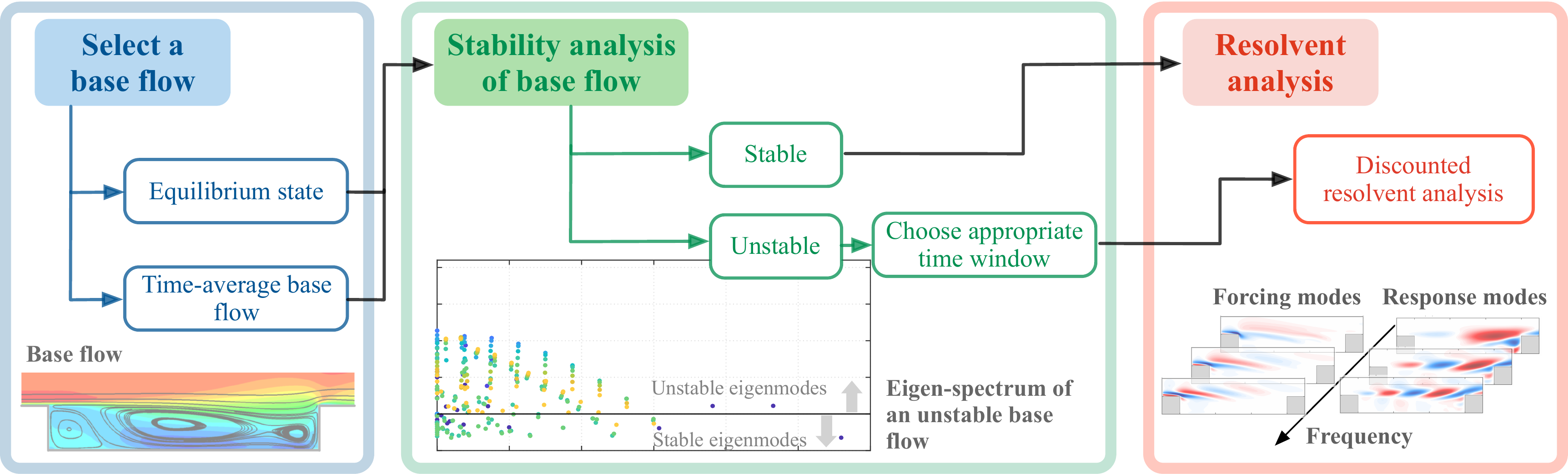}
\caption{Overview of the resolvent analysis.}
\label{fig:overview}
\end{figure}

Let us present the resolvent analysis approach necessary to examine the input-output properties of fluid flows.  The overall approach is illustrated in figure \ref{fig:overview} that we will discuss each component of the resolvent analysis below.  

\subsubsection{Resolvent operator}
\label{sec:method_resl}
Resolvent analysis is an operator-based modal analysis formulation that is capable of examining flow response to harmonic forcing input with respect to a given base state. For this analysis, the base state can be an equilibrium state or a time-average flow, provided that the flow is in statistical equilibrium. The dimensionless governing equation for fluid flow is the (spatially discretized) Navier--Stokes equation expressed as 
\begin{equation}
\label{eqn:NS}
\frac{\partial {\boldsymbol q}}{\partial t} = N(\boldsymbol q),
\end{equation}
where the state variable $\boldsymbol q = [\rho,  \rho u, \rho v, \rho w, e]^T \in \mathbb R^{5m}$.  Here, we have density $\rho$, velocity $u$, $v$, and $w$ for streamwise $x$, transverse $y$, and spanwise $z$ directions, and energy $e$.  The number of grid points used to discretize the computational domain is denoted by $m$.  All variables have been non-dimensionalized by the cavity depth and free stream values of the variables, and velocity is non-dimensionalized by free stream acoustic speed as described in \cite{Sun:JFM17}. We decompose the flow state $\boldsymbol q$ into the 2D steady base state $\boldsymbol {\bar q}$ and the 3D perturbation $\boldsymbol {q'}$
\begin{equation}
\label{Re_decmp}
\boldsymbol q(x,y,z,t) = \boldsymbol {\bar q}(x,y) + \boldsymbol {q}'(x,y,z,t).
\end{equation}
Substituting this expression into Eq.~(\ref{eqn:NS}), the governing equation for the perturbation becomes
\begin{equation}\label{eqn:decom_NS}
\frac{\partial \boldsymbol q'}{\partial t}= \tilde L(\boldsymbol{\bar q})\boldsymbol q'+   \boldsymbol f',
\end{equation}
where $\tilde L(\boldsymbol {\bar q})$ denotes the linear operator of the Navier--Stokes equations on $\boldsymbol q'$ and the term $\boldsymbol f'$ represents a forcing applied to the linear dynamical system. The above linearization holds only if $\boldsymbol {\bar q}$ is an equilibrium state. For non-equilibrium $\boldsymbol {\bar q}$, the forcing term $\boldsymbol f'$ can be interpreted as a combination of the right-hand-side of the original Navier--Stokes equation with respect to the base state, the nonlinear higher-order perturbation terms, and/or external forcing introduced into the fluid flow system.   With the assumption that the flow to be examined is in a statistically stationary state, both forcing $\boldsymbol f'$ and perturbation $\boldsymbol q'$ can take the Fourier representations of  
\begin{equation}
\boldsymbol f' (x,y,z,t)= \boldsymbol {\hat{f}}_{\omega,\beta}(x,y)e^{i(\beta z-\omega t)} + \text{complex conjugate},
\label{modal_f}
\end{equation}
\begin{equation}
\boldsymbol q' (x,y,z,t)= \boldsymbol {\hat{q}}_{\omega,\beta}(x,y)e^{i(\beta z-\omega t)} + \text{complex conjugate},
\label{modal_q}
\end{equation}
with real-valued radian frequency $\omega$, real-valued spanwise wavenumber $\beta$, and $\boldsymbol {\hat{f}}_{\omega,\beta}$ and $\boldsymbol {\hat{q}}_{\omega,\beta}$ denoting spatial amplitude functions for forcing and perturbation, respectively. Here, we have assumed that the flow is spanwise periodic, although such assumption can be removed in general.  By substituting Eqs.~(\ref{modal_f}) and (\ref{modal_q}) into Eq.~({\ref{eqn:decom_NS}}), the governing equation can be rewritten as 
\begin{equation}
-i\omega \boldsymbol{\hat q}_{\omega,\beta} =   L (\boldsymbol{\bar q};\beta)\boldsymbol{\hat q}_{\omega,\beta} + \boldsymbol{\hat f}_{\omega,\beta}.
\label{resl_ver1}
\end{equation}

When the system is unforced ($\boldsymbol f'=0$), Eq.~(\ref{eqn:decom_NS}) reduces to an initial value problem, which can be solved for the homogeneous solution of Eq.~(\ref{resl_ver1}) with $\boldsymbol{\hat f}_{\omega,\beta}=0$.  In other words, this can be formulated as an eigenvalue problem that analyzes the instabilities of the base flow.  When the forcing term is nonzero, Eq.~(\ref{resl_ver1}) describes the relationship between the flow response and harmonic forcing, which provides the particular solution of the inhomogeneous equation (Eq.~(\ref{resl_ver1})) for the system with a sustained forcing input. 

For the forced system, by solving for $\boldsymbol {\hat{q}}_{\omega,\beta}$ from Eq.~(\ref{resl_ver1}), we find 
\begin{equation}\label{eqn:Resolvent}
\boldsymbol {\hat q}_{\omega,\beta} = [-i\omega   I -   L(\boldsymbol {\bar q}; \beta)]^{-1}\boldsymbol{\hat f}_{\omega,\beta}=  {H}({\boldsymbol{\bar q};\omega, \beta})\boldsymbol {\hat f}_{\omega,\beta},
\end{equation}
where $H({\boldsymbol{\bar q}; \omega, \beta})=[-i\omega   I -   L(\boldsymbol{\bar q}, \beta)]^{-1}$ is referred to as the {\it resolvent operator}. Here, $H({\boldsymbol {\bar q}; \omega, \beta})$ is the transfer function between the input $\boldsymbol {\hat f}_{\omega,\beta}$ and the output $\boldsymbol{\hat q}_{\omega,\beta}$ for the given base state $\boldsymbol {\bar q}(x,y)$, real-valued spanwise wavenumber $\beta$ and real-valued frequency $\omega$ \cite{Astrom:2010}. The selection of the base state is important as discussed below.

\subsubsection{Choice of base state}
The base state chosen for resolvent analysis should be a time-invariant state, such as the equilibrium or the time-average (mean) flow.  For flows up to moderate Reynolds numbers, the equilibrium state can be solved for via the selective frequency damping method or a Newton--Krylov-type iterative method \cite{Akervik:PF06,Tuckerman:IMA00, Kelley:arxiv}. 
However, for turbulent flows at higher Reynolds numbers, solving for the equilibrium state can be difficult.  Moreover, the existence of the equilibrium point itself may be questionable.  In such a case, the statistically stationary mean flow can be used as the base state for constructing the linear operator $L(\boldsymbol {\bar q};\beta)$.  Although stability analysis is valid only when the base state $\boldsymbol {\bar q}$ is an equilibrium state of the flow, the exercise of checking instability of $L(\boldsymbol {\bar q};\beta)$ based on the mean flow is a precursor for performing resolvent analysis to seek proper physical interpretation. 

\subsubsection{Stability analysis}
\label{sec:method_GSA}

Given a stable base state, the interpretation of the findings from resolvent analysis is straightforward.  The response of the forcing input is amplified by the respective gain amplitude with the spatial structure of the response mode.  The sustained periodic forcing input yields periodic response output.  However, the results from resolvent analysis should be taken with care if the given base state is unstable.  If the base state is unstable, perturbations added to the base flow can grow larger than the harmonic output, leaving the insights from resolvent analysis in vein.  To obtain proper insights from resolvent analysis for unstable base flows, the instabilities of the base state $\boldsymbol{\bar q}$ should be taken into consideration.  Stability analysis examines development of small perturbations about a base state \cite{Theofilis:ARFM11} via performing eigen-decomposition of the linearized Navier--Stokes operator $L(\boldsymbol {\bar q};\beta)$.  Considering the unforced system ($\boldsymbol{\hat f}=0$) with
Eq.~(\ref{resl_ver1}), we form a temporal eigenvalue problem shown as 
\begin{equation}
-i\tilde \omega \boldsymbol{\hat q}_\beta =  {L(\boldsymbol {\bar q};\beta)}\boldsymbol{\hat q}_\beta,
\label{modal2}
\end{equation}
where eigenvalue $-i\tilde \omega$ is a complex-valued number with real ($\tilde \omega_r$) and imaginary ($\tilde \omega_i$) components representing the temporal frequency and growth ($\tilde \omega_i>0$) or decay ($\tilde \omega_i<0$) rate, respectively, of the instability.  Strictly speaking, stability analysis should be performed about an equilibrium state that is a solution to the Navier--Stokes equations.  Here, the stability analysis for a time-average base state is not focused on the identification of instabilities but is utilized to determine the temporal window that must be used in the resolvent analysis, as discussed later. 

If the linear operator contains only stable eigenmodes such that $\max(\tilde \omega_i)<0$, we can directly use the formulation shown in section \ref{sec:method_resl} to perform resolvent analysis and examine the asymptotic behaviors of the dynamical system in an infinite-time horizon.  If the linear operator possesses unstable eigenmodes such that $\max(\tilde \omega_i)\ge0$, the resolvent analysis should be modified to have a time-window characterized by a real-valued parameter $\alpha$ satisfying $\alpha>\max(\tilde \omega_i)$.  This leads to the so-called {\it discounted resolvent analysis} proposed by Jovanovi\'c \cite{Jovanovic:04}, which examines the harmonic response of the dynamical system within the finite-time window before instabilities grow unboundedly.  With these subtle but important points in mind, we introduce the resolvent analysis.

\subsubsection{Resolvent analysis}

For a stable base state, the resolvent analysis can be cast in the framework of singular value decomposition (SVD) of the resolvent operator to determine the dominant forcing directions $\boldsymbol{\hat f}_{\omega,\beta}$ and the output directions $\boldsymbol{\hat q}_{\omega,\beta}$.  Using the matrix notation, we can express the decomposition as
\begin{equation}\label{eqn:svd}
	H({\boldsymbol{\bar q}; \omega, \beta}) =  \mathcal Q \Sigma  \mathcal F^\ast ,
\end{equation}
where $H\in \mathbb{C}^{5m\times5m}$, $\mathcal Q=[\boldsymbol {\hat q}_1, \boldsymbol{\hat q}_2,\dots, \boldsymbol{\hat q}_k]$ contains a set of left singular vectors $\boldsymbol {\hat q}_j$ called the {\it response modes}, and $\mathcal F=[\boldsymbol{\hat f}_1, \boldsymbol{\hat f}_2,\dots, \boldsymbol{\hat f}_k]$ holds a set of right singular vectors $\boldsymbol {\hat f}_j$ called the {\it forcing modes} with the superscript $\ast$ representing the Hermitian transpose.   Amplification ratio of response and forcing modes are given by the corresponding singular values $\Sigma=\text{diag}(\sigma_1,\sigma_2,\dots,\sigma_k)$ in descending order, where the leading amplification $\sigma_1$ is referred to as the {\it resolvent gain}.  Alternatively Eq.~(\ref{eqn:svd}) can be viewed as $\mathcal Q \Sigma = H \mathcal F$, where the resolvent operator maps the unit vector $\boldsymbol {\hat f}_j$ to the unit vector $\boldsymbol {\hat q}_j$ with the gain $\sigma_j$. 

The resolvent gain is studied in the context of having an energy norm defined by $E=\int _S  (|\hat \rho|^2 +|\hat {\rho u}|^2+|\hat {\rho v}|^2+|\hat {\rho w}|^2+|\hat e|^2) dA = ||\boldsymbol {\hat q}||_E^2 $ in the present work. This norm can be related to the induced 2-norm of the resolvent operator $H({\boldsymbol{\bar q}; \omega, \beta})$ through a weight matrix $W$, such that $||\boldsymbol {\hat q}||_E^2 = || W\boldsymbol{\hat q}||_2^2$ \cite{Schmid01,Yeh:JFM19}. The weight matrix $W$ can be constructed based on the discretization scheme adopted in the numerical configuration. Consequently, the optimal ratio of this energy norm of response to forcing modes can be obtained via calculating the largest singular value of the resolvent operator, which is the induced 2-norm of the weighted resolvent matrix $WH({\boldsymbol{\bar q}; \omega, \beta})W^{-1}$. Because of the weight matrix $W$ used in the induced 2-norm evaluation, the resulting forcing and response modes shown later have been scaled by $W^{-1}$ to represent the correct flow field. Details on the norms can be found in Yeh \& Taira \cite{Yeh:JFM19}. 

\subsubsection{Discounted resolvent operator}

In discounted resolvent analysis \cite{Jovanovic:04} for an unstable flow, an exponential discount is introduced to the dynamical system, in which energy amplification within a finite-time window can be examined. The finite-time window is realized by introducing a real-valued parameter $\alpha$ into the original resolvent operator (Eq.~(\ref{eqn:Resolvent})) to form the discounted resolvent operator
\begin{equation}\label{eqn:dis}
H^\alpha({\boldsymbol {\bar q}; \omega, \beta})=[- i(\omega+\text{i} \alpha)   I-   L(\boldsymbol{\bar q}; \beta)]^{-1} = [- i\omega I - L(\boldsymbol{\bar q}; \beta) + \alpha I]^{-1}.
\end{equation}
The eigenspectrum of maxtrix $[L(\boldsymbol {\bar q};\beta)-\alpha I]$ is equivalent to the eigenspectrum of $L(\boldsymbol {\bar q};\beta)$ shifted by $-\alpha$ along imaginary axis. In other words, the flow response under investigation is in a finite-time window characterized by $1/\alpha$. As $\alpha \to 0^+$, the window becomes infinite such that Eq.~(\ref{eqn:dis}) reduces to the original resolvent operator in Eq.~(\ref{eqn:Resolvent}). For an unstable flow with $\max(\tilde \omega_i)>0$, by choosing $\alpha>\max(\tilde \omega_i)$, we select to examine the harmonic response of the flow prior to the instabilities invalidating the formulation.  It is important to note that the actual flow will not grow unboundedly.  Nonlinearity will act to saturate the disturbance level, and its magnitude will stay bounded.  In many cases, including flow control applications, the resolvent analysis is hoped to provide the dominant amplification that likely is going to trigger nonlinear effects to shift the mean flow to a desirable state for achieving some engineering benefits.

\subsection{Compressible laminar and turbulent open-cavity flows}

In the present work, we examine laminar and turbulent compressible flow over a long rectangular cavity.  The cavity of length-to-depth ratio $L/D=6$ and width-to-depth ratio $\Lambda/D=2$ is considered with the free stream having a Mach number of $M_\infty=0.6$. The compressible flow solver {\it CharLES} \cite{Khalighi:ASME2011,Khalighi:AIAA11,Bres:AIAAJ17} is used to perform direct numerical simulation and large eddy simulation for the cavity flow at depth-based Reynolds numbers of $Re_D=502$ and $10^4$, respectively.  A second-order finite-volume method and third-order Range--Kutta temporal integration scheme are used.  Given the initial momentum thickness $\theta_0$ at the cavity leading edge, the corresponding momentum-thickness-based Reynolds numbers are $Re_\theta=19$ and $225$, respectively.  

\begin{figure}[hbpt]
\centering
\includegraphics[width=0.5\textwidth]{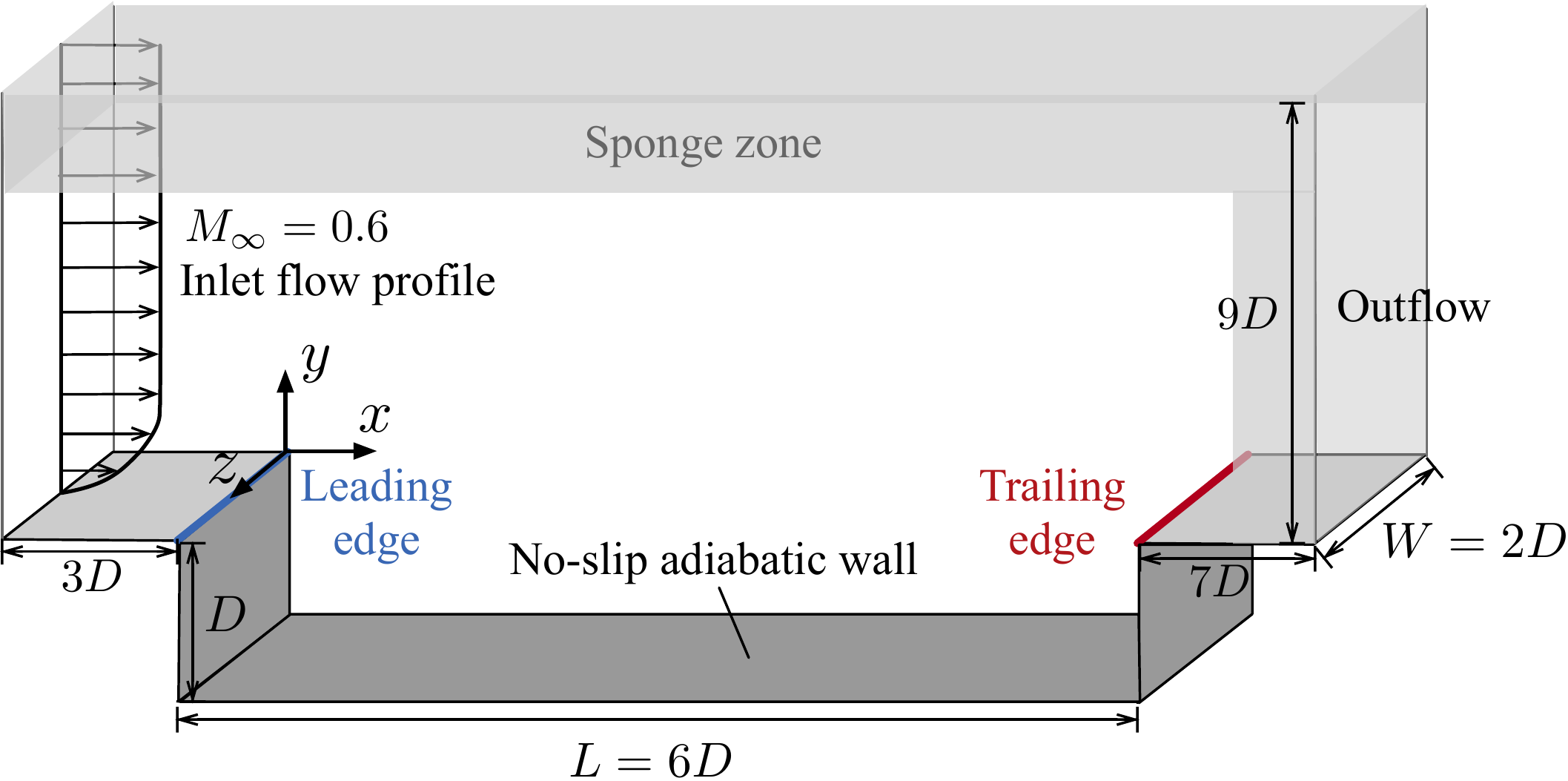}
\caption{The computational setup for flow over a rectangular cavity with $L/D=6$ and $W/D=2$ (not to scale).}
\label{fig:setup}
\end{figure}

The computational setup is shown in figure \ref{fig:setup}. The origin of Cartesian coordinate system is placed at the cavity leading edge with $x$, $y$ and $z$ representing the streamwise, transverse and spanwise directions, respectively, with corresponding velocity components denoted by $u$, $v$ and $w$. No-slip and adiabatic boundary conditions are prescribed at the cavity walls. Spanwise periodicity is prescribed with cavity span of $W=2D$. We specify a laminar Blasius boundary layer profile for incoming flow at $Re_D=502$ with an initial boundary layer thickness of $D/\delta_0=3.6$ at the cavity leading edge. For the turbulent flow at $Re_D=10^4$, we use a turbulent mean velocity profile using the one seventh power law, and superpose random Fourier modes to approximate a turbulent boundary layer profile \cite{Bechara:AIAAJ94,Franck:AIAAJ10}. The initial boundary layer thickness is set to $D/\delta_0 = 6$. The inlet boundary is placed 3D from the cavity leading edge, and the outflow boundary is placed 7D from the cavity trailing edge. The far-field boundary condition is placed 9D above the cavity leading edge. Sponge zones spanning $2D$ are applied at outlet and far-field boundaries to damp out exiting waves and prevent reflections. Structured meshes with 7 and 14 million grid points are used for cases with $Re_D=502$ and $10^4$, respectively. The grid resolutions have been found to be sufficient to resolve the flow structures as discussed in our previous studies \cite{Sun:TCFD16,Sun:AIAAJ19}. 

In the stability and resolvent analyses, the Dirichlet boundary condition at the inlet for velocity and pressure gradient of the perturbation variables are specified to be zero. For outflow and far-field boundaries, Neumann boundary conditions for density, velocity and pressure are prescribed as zero. Along the cavity walls, velocity and transverse gradient of pressure perturbations are set to zero. Since the size of the linear operator ${L(\boldsymbol {\bar q};\beta)}\in \mathbb{C}^{5m \times 5m}$, where $5m$ can be extremely large due to number of grid points used to discretize the flow, we use ARPACK library \cite{Arpack:96} to solve the large-scale eigenvalue problem. To expedite the eigen-decomposition of the linear operator $L(\boldsymbol {\bar q};\beta)$, we have interpolated the mean flow onto a coarse mesh with 0.8 million points \cite{Bergamo:AST15}, for which the domain size and grid resolution have also been examined to be sufficient to capture the dominant eigenmodes \cite{Sun:TCFD16,Sun:JFM17}.

\section{Results and Discussions}
\label{sec:result}

In this section, we examine the characteristics of rectangular cavity flows at $M_\infty=0.6$ and Reynolds numbers of $Re_D=502$ and $10^4$. For both cases, stability and resolvent analyses are performed to uncover the underlying 2D and 3D flow physics.

\subsection{Baseline flow characteristics}
\label{sec:base}

\begin{figure}[hbpt]
\includegraphics[width=0.98\textwidth]{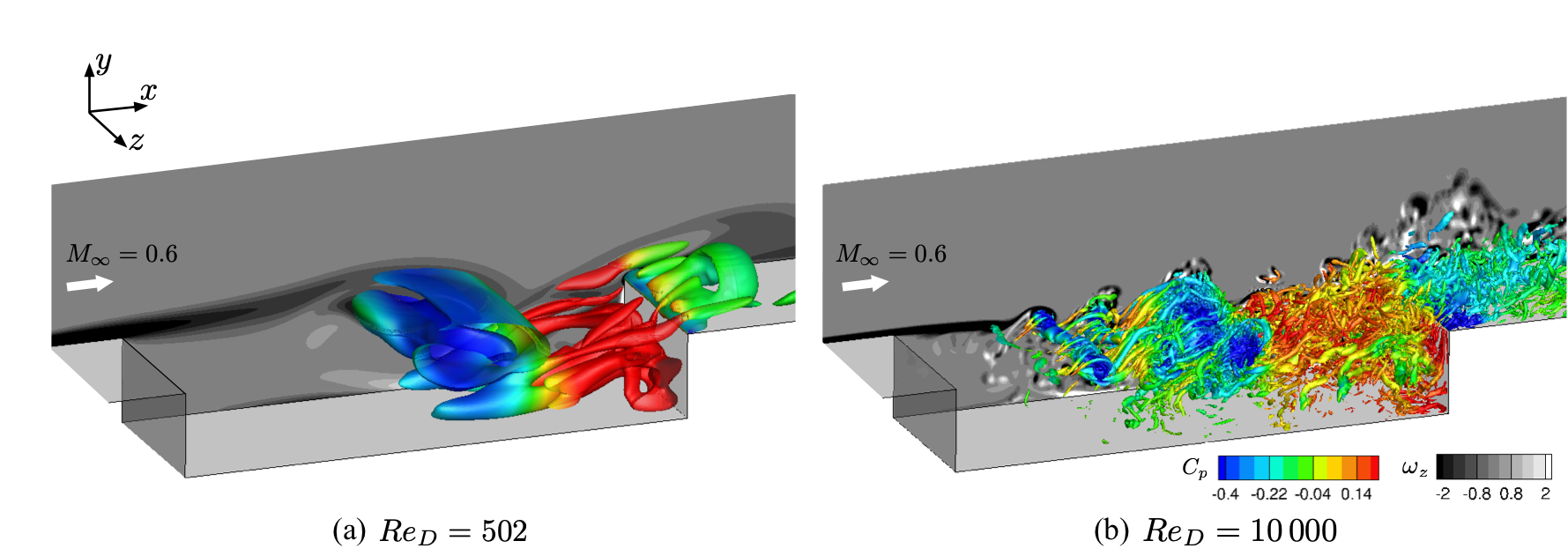}
\caption{Iso-surfaces of $Q$-criterion colored by the pressure coefficient $C_p=(p-p_\text{ref})/(\frac{1}{2}\rho u_\infty^2)$ of instantaneous flow fields at (a) $Re_D=502$ with $Q(D/u_\infty)^2=0.33$ and (b) $Re_D=10^4$ with $Q(D/u_\infty)^2=14$. }
\label{fig:instant}
\end{figure}

The primary feature of open-cavity flow is the roll up of the shear layer emanating from the cavity leading edge, which impinges on the cavity aft-wall and generates large pressure fluctuations. Representative instantaneous flow fields for the two different Reynolds numbers are shown in figure \ref{fig:instant}. The iso-surfaces show the $Q$-criterion \cite{Hunt:88} highlighting the coherent vortical structures. The background contour visualizes the spanwise vorticity, which exhibits roll-up of the shear layer as the flow passes the leading edge. At $Re_D=502$, a large vortex appears at the center of the cavity ($x/D \approx 3$), and its structure exhibits low-level of spanwise variation. However, below the coherent structures, streamwise vortices are generated inside the cavity and develop towards the cavity aft-wall.  For the turbulent flow at $Re_D=10^4$, the spanwise coherent structures form farther upstream than the laminar flow. These structures gradually become larger as they merge with other structures. For both cases, the most intense pressure fluctuations are experienced at the cavity aft-walls.  While there are differences in the detailed flow structures, the common mechanism of shear-layer instabilities over the cavity remains the same regardless of the Reynolds number.

\begin{figure}[hbpt]
\includegraphics[width=0.98\textwidth]{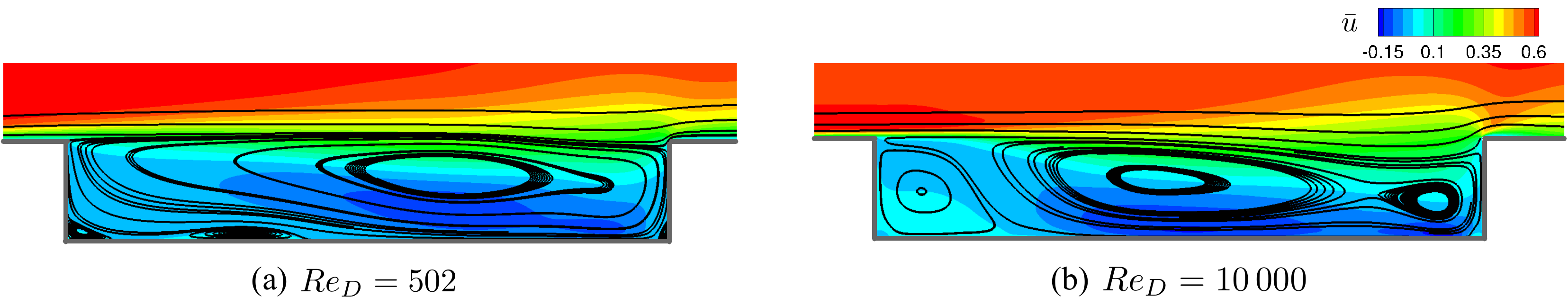}
\caption{Streamlines with streamwise velocity $\bar u$ contour in the background for time- and spanwise-average flows at (a) $Re_D=502$ and (b) $Re_D=10^4$.}
\label{fig:mean}
\end{figure}

The time- and spanwise-average mean flow, which serves as the base state for the present stability and resolvent analyses, is shown in figure \ref{fig:mean}. For laminar and turbulent flows, the overall feature of the mean flow remains similar especially in the shear layer region although the streamlines of the turbulent flow near the trailing edge penetrate deeper into the cavity.  For laminar flow, a recirculation region covers the entire streamwise extent of the inside cavity with its center located around $x/D\approx4$. Three small recirculation zones are observed on cavity floor with negligible size compared to the primary one. For turbulent flow, three sizable recirculation zones are present inside the cavity. The center of the primary recirculation zone is positioned around the center of the cavity. 

\begin{figure}[hbpt]
\centering
\includegraphics[width=0.95\textwidth]{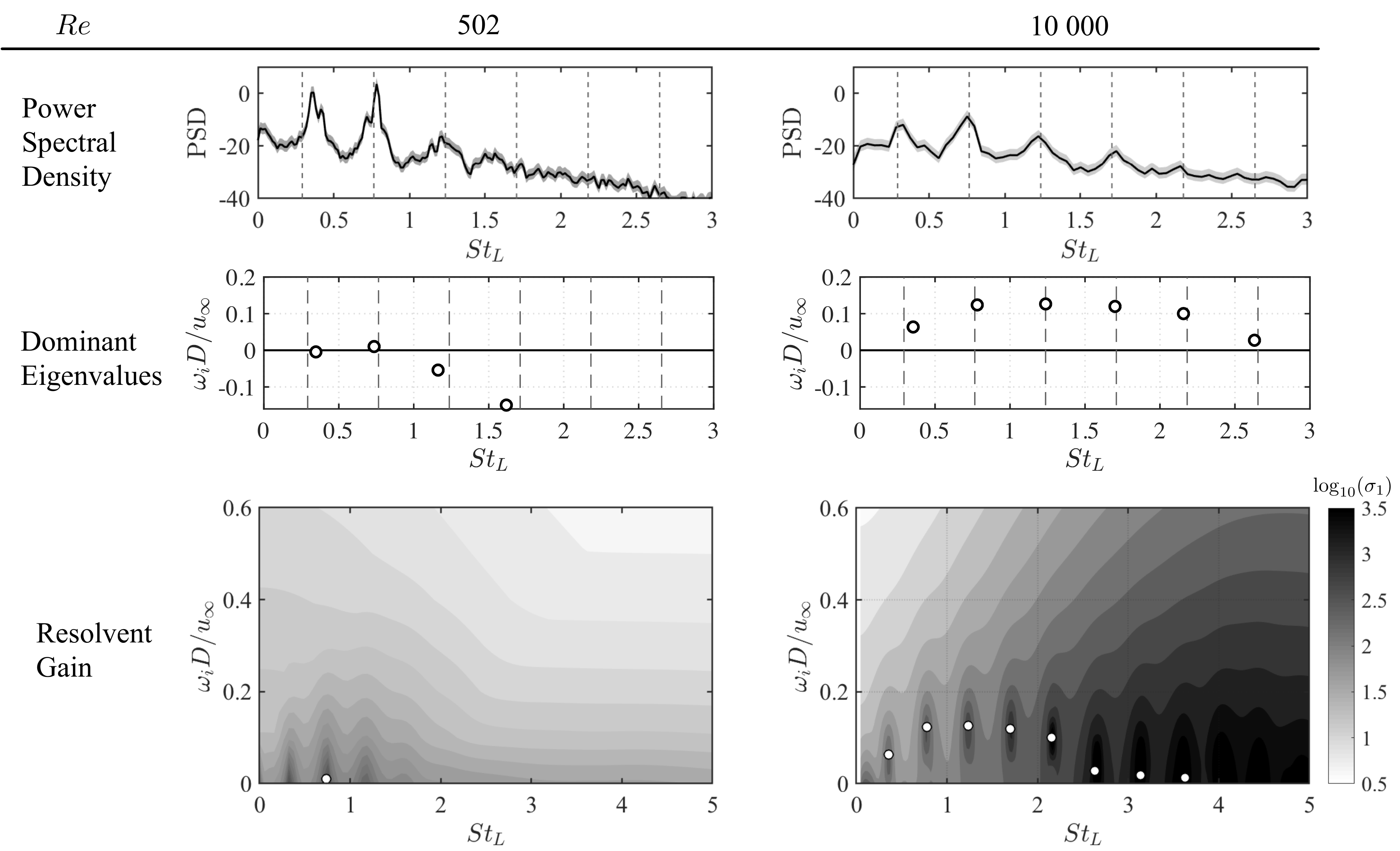}
\caption{(Top) power spectral density of aft-wall pressure at $[x, y, z]/D = [6, -0.5, 0]$ from nonlinear simulations. (Middle) dominant eigenvalues from stability analysis with Rossiter mode frequencies (dashed lines) predicted from the semi-empirical formula. (Bottom) pseudospectra (discounted resolvent gain) of the linear operator $L(\boldsymbol {\bar q};\beta)$ with $\beta = 0$, where the dominant eigenvalues are also displayed.}
\label{fig:2Dfreq}
\end{figure}

A prominent feature in open-cavity flow is the presence of strong resonant tones. The underlying physics of these tones will be examined using the stability and resolvent analysis discussed later. The resonant modes in open-cavity flow are called {\it Rossiter modes} \cite{Rossiter:ARCRM64} whose frequencies can be calculated using a semi-empirical formula \cite{heller1975physical} given by
\begin{equation}
St_L=\frac{f_nL}{u_\infty}=\frac{n-\tilde \alpha}{1/\kappa+M_\infty/\sqrt{1+(\gamma-1)M_\infty^2/2}},
\label{eq:Rossiter}
\end{equation}
where $f_n$ is the frequency of $n$th Rossiter mode, $\tilde \alpha=0.38$ is the phase delay, $\kappa=0.65$ is an empirical constant representing the average convective speed of disturbance in the shear layer \cite{Rossiter:ARCRM64}, and $\gamma=1.4$ is specific heat ratio.  To identify the Rossiter modes, power spectral density plots of the pressure history are shown in figure \ref{fig:2Dfreq} using Welch's method with Hanning window and 75\% overlap. The pressure is normalized by the free stream dynamic pressure, and the gray shaded areas indicate uncertainty bounds representing 95\% confidence. The peaks of power spectral density agree well with the predicted Rossiter mode frequency in both low and high Reynolds number cases. As observed in figure \ref{fig:instant}, the spanwise coherent structure is larger in size at $Re_D=502$ than at $Re_D=10^4$. For that reason, the strength of its impingement on the aft-wall becomes more violent such that the dominant Rossiter modes (I and II) are stronger for the low Reynolds number flow. In contrast, turbulent flow has more small-scale structures resulting in the pressure fluctuation broadband especially in the high frequency region of $St_L>2$.

\subsection{Two-dimensional instability and resolvent modes}
\label{sec:2D}

Although we have seen highly three-dimensional structures in the instantaneous flow, the shear-layer instability is indeed a 2D mechanism driving the flow oscillations. Hence, we perform stability and resolvent analyses using time- and spanwise-average mean flow as the base flow and set spanwise wavenumber $\beta$ to zero to extract 2D modes.  The eigenvalues of the 2D eigenmodes obtained from stability analysis are shown in figure \ref{fig:2Dfreq} over $St_L=\omega_rL/(2\pi u_\infty)$ and $\omega_iD/u_\infty$ representing Strouhal number and growth/decay rate, respectively. The frequencies of the 2D eigenmodes agree well with the power spectral density. This observation has also been noted in other studies \cite{Sipp:JFM07,Turton:PRE15,Sun:TCFD16} that a use of mean flow as base state in stability analysis yields good prediction of temporal frequencies present in the unsteady flows. While the large value of growth rate $\tilde \omega_iD/u_\infty$ in $Re_D=502$ case can be correlated with the dominance of Rossiter mode II, this relation is not clear for the $Re_D=10^4$ case.

Because stability analysis for the laminar and turbulent flows captures unstable eigenmodes, the discounted resolvent analysis with $\alpha>\max(\tilde \omega_i)$ needs to be considered. As the first singular value $\sigma_1$ of resolvent operator is assessed using the pseudospectra of the linear operator $L(\boldsymbol{\bar q}; \beta)$ along the real axis \cite{Trefethen:05}.  Here, we present the pseudospectrum of the linear operator $L(\boldsymbol{\bar q};\beta)$ in figure \ref{fig:2Dfreq}.  The gain of discounted resolvent analysis can be determined also for the finite-time window characterized by evaluating the pseudospectra along $1/\alpha = u_\infty/(\omega_i D)$. The dominant eigenvalues obtained from stability analysis are also displayed for reference. Both laminar and turbulent flows exhibit large resolvent gain in the vicinity of the eigenvalues with large amplification of the energy norm.  For the turbulent flow, additional areas of higher gain appear at frequency $St_L\gtrsim4.5$ away from the eigenvalues. We note that if a linear operator is highly non-normal ($L^*L\ne LL^*$), the dynamical response does not solely dependent on the characteristics of individual eigenmode, as regions of high gain can extend far away from the location of the eigenvalues \cite{Trefethen578}.  That is, the pseudospectra can exhibit slow decay of the gain away from the poles (eigenvalues).  In the present study, the extra high-gain area in the turbulent flow case results from non-normality of the linear operator, which is not observed in the laminar flow.  Moreover, the frequency associated with the highest resolvent gain is far away from the dominant resonance in the nonlinear flow, which suggests that the high-gain region might not exactly correspond to the primary instabilities in the flow revealed from stability analysis, but it could potentially provide insights on how to excite flows efficiently with respect to control-oriented studies.  In the discussions here, we only focus on the dominant modes but the analysis can reveal subdominant modes as well.

\subsection{Three-dimensional instability and resolvent modes}
\label{sec:3D_gsa}

\begin{figure}[hbpt]
	\begin{center}
		\includegraphics[width=0.99\textwidth]{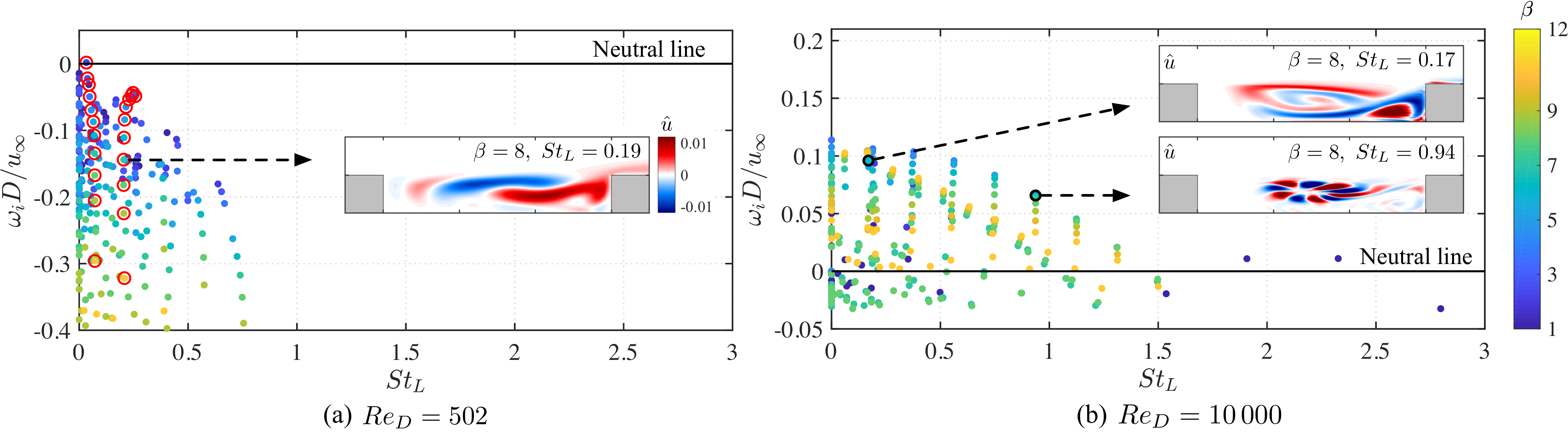}
	\caption{Eigenspectra of 3D eigenmodes with $1 \le \beta \le 12$ for $Re_D=$ (a) 502 and (b) $10^4$. The two most dominant eigenmodes for the laminar flow case are indicated by the red circles. Representative eigenmodes with $\beta=8$ are visualized using the real component of streamwise velocity $\hat u$.}
	\label{fig:3D_eig}
	\end{center}
\end{figure}

Let us discuss the three-dimensional instabilities with spanwise wavenumber $\beta>0$. Shown in figure \ref{fig:3D_eig} are the eigenspectra and selective eigenvectors of the laminar and turbulent cavity flows with spanwise wavenumber over a range of $1\le \beta \le 12$. The maximum value of growth rate $\tilde \omega_i D/u_\infty$ will guide us to choose the proper value of $\alpha$ for the discounted resolvent analysis to be performed later. The eigenspectra associated with each $\beta$ are denoted by different colors. 

For the laminar flow (figure \ref{fig:3D_eig} (a)), the eigenspectra are concentrated in the low frequency region with $St_L<1$. As $\beta$ increases, the eigenmodes become more stable, and the eigenvalues exhibit a spreading pattern in terms of modal frequency. Among these eigenmodes, the least-stable ones are associated with low wavenumbers and low frequencies, which have been reported in previous studies \cite{Sun:TCFD16}. These low wavenumber modes stem from centrifugal instabilities that are collocated with the recirculation region within the cavity. As shown in the subplot of the eigenvector of the leading eigenmode with $\beta=8$ in figure \ref{fig:3D_eig} (a), the disturbance of streamwise velocity resides around the primary recirculation and fluctuates with a low frequency of $St_L=0.19$. 

For the turbulent flow (figure \ref{fig:3D_eig} (b)), unstable eigenmodes are captured covering a wider range of frequencies than that of the laminar flow. The most-unstable mode has a  wavenumber around $\beta=5$, which is different from the laminar flow case where the least-stable mode has a wavenumber of $\beta=1$. For the turbulent case, the eigenspectra for $\beta=1$ do not follow the pattern presented by the large wavenumber cases with $\beta \ge 2$, and the frequencies of the eigenmodes ($\beta=1$) are higher than those of the eigenmodes with $\beta \ge 2$. Moreover, centrifugal instabilities with low frequencies are also observed in the turbulent flow, but there is a new branch of eigenmodes with high frequencies as illustrated in subplots of figure \ref{fig:3D_eig} (b). The disturbance in these eigenmodes mainly oscillates in the area of the center recirculation zone as shown in the mean flow profile (figure \ref{fig:mean}).

\begin{figure}[hbpt]
	\begin{center}
		\includegraphics[width=0.98\textwidth]{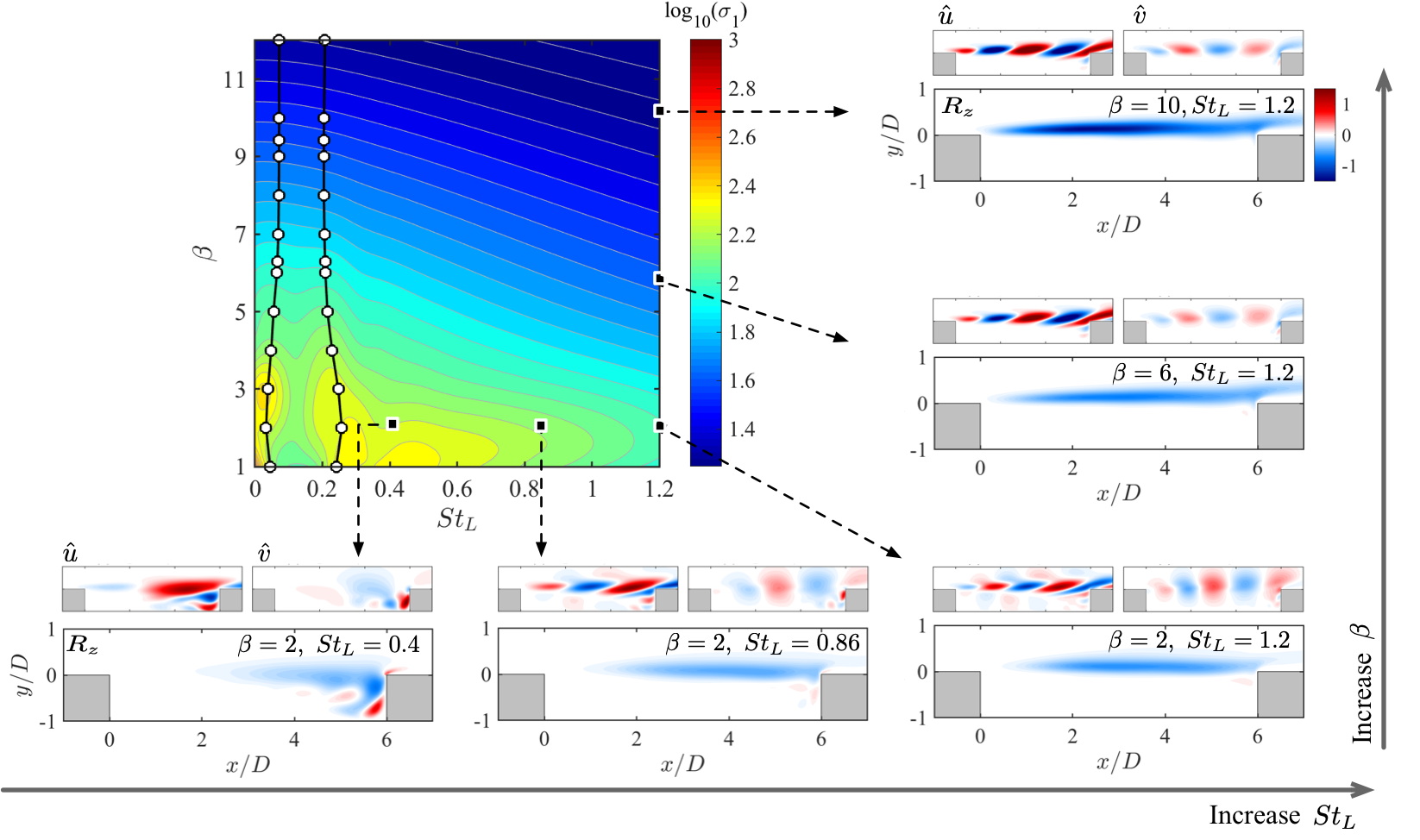}
	\caption{The leading gains displayed as contours from resolvent analysis with $\alpha=0$, and leading eigenvalues of corresponding $\beta$ from stability analysis are denoted by white circles at $Re_D=502$. The white dots correspond to the two most dominant eigenmodes shown in figure \ref{fig:3D_eig} (a). Subplots select first response modes visualized using the real component of streamwise velocity $\hat u$, transverse velocity $\hat v$ and induced Reynolds stress $R_z$. }
	\label{fig:3D_gain_low}
	\end{center}
\end{figure}

With the instabilities of the base state known, we can now perform resolvent analysis with a proper choice of $\alpha$ for the discounted resolvent analysis. For the laminar flow, since all eigenvalues are in the stable plane ($\tilde \omega_iD/u_\infty<0$) (figure \ref{fig:3D_eig}), we perform the resolvent analysis using the standard resolvent operator Eq.~(\ref{eqn:Resolvent}) with $\alpha=0$. The 3D resolvent modes and the associated gains with a range of $1 \le \beta \le 12$ are shown in figure \ref{fig:3D_gain_low}. The variation of the two leading eigenvalues from figure \ref{fig:3D_eig} (denoted by red circles) are superposed on the gain map, for which the highest resolvent gains appear around low spanwise wavenumber of $\beta \approx 3$ and low frequency with $St_L\lessapprox 0.3$, indicating that the forcing associated with these values of $\beta$ and $St_L$ will be amplified the most in terms of the energy norm. Moreover, for increasing $\beta$ beyond 4 and increasing $St_L$ beyond 0.3, the gain decreases monotonically for the considered range.

Local peaks in the resolvent gain (sliced horizontally) are observed for $\beta < 7$ of the laminar flow. The eigenvalues from stability analysis reside at these high gain peaks. This overlap in frequencies of the most responsive mode and eigenmode from resolvent and stability analyses can be explained from the eigenspectrum shown in figure \ref{fig:3D_eig} (a). Because the resolvent gain here is equivalent to the pseudospectrum along the neutral stability line of $\omega_iD/u_\infty = 0$, the optimal gain distribution is influenced by the eigenmodes in the vicinity of the neutral stability line. As the two leading eigenmodes are the nearest modes below the neutral stability line, two local maxima are revealed from the gain distribution along the frequency axis. We also note that the normality of a matrix can determine the gain behavior \cite{Trefethen578, Schmid01}. The shown cases yield a normal linear operator, for which the frequencies of gain peaks and eigenmodes match, and the gain value can be well approximated by the inverse of the distance between the eigenvalue and neutral stability line. This interpretation explains the appearance of maximum gain in relation to the locations of eigenvalues.  For the range of $\beta \ge 7$, the gain decreases as frequency increases, and there are no distinct peaks appearing at the frequencies of the leading eigenmodes.  The reason for this is due to the eigenmodes with $\beta>7$ being quite stable with their eigenvalues placed far away from the neutral stability line. 

Next, let us examine the influence of spanwise wavenumber $\beta$ and frequency $St_L$ on the features of the response mode. Response modes are plotted along $\beta=2$ and $St_L=1.2$ in figure \ref{fig:3D_gain_low}.  Among the response modes of $\beta=2$, for $St_L$ increases from 0.4 to 1.2 (beyond the frequency of the centrifugal modes), the scale of the spatial structure becomes smaller, and the response modes are concentrated in the shear-layer region, which are similar to the shear-layer instabilities from stability analysis for $\beta=0$. For the response modes of $St_L=1.2$, when $\beta$ increases from 2 to 10, the most distinct changes in the modal structures are the reduction in fluctuations over the lower and upper regions of the shear layer, and enhanced fluctuations in the shear-layer region. Moreover, the location of the highest magnitude of $\hat u$ and $\hat v$ components moves upstream from the latter part of the cavity shear layer near the trailing edge to the leading edge. Here, we further calculate the spanwise Reynolds stress using the $\hat u$ and $\hat v$ components of the response mode 
\begin{equation}
R_z=\Re(\hat u^*\hat v),
\end{equation}
where $^*$ represents complex conjugate, and $\Re(\cdot)$ indicates the real component. High distribution of the Reynolds stress appears upstream in the shear layer for high-frequency response modes, indicating that mixing in shear layer occurs immediately aft of the cavity leading edge for large spanwise wavenumber $\beta$.  

\begin{figure}[hbpt]
	\begin{center}
		\includegraphics[width=0.35\textwidth]{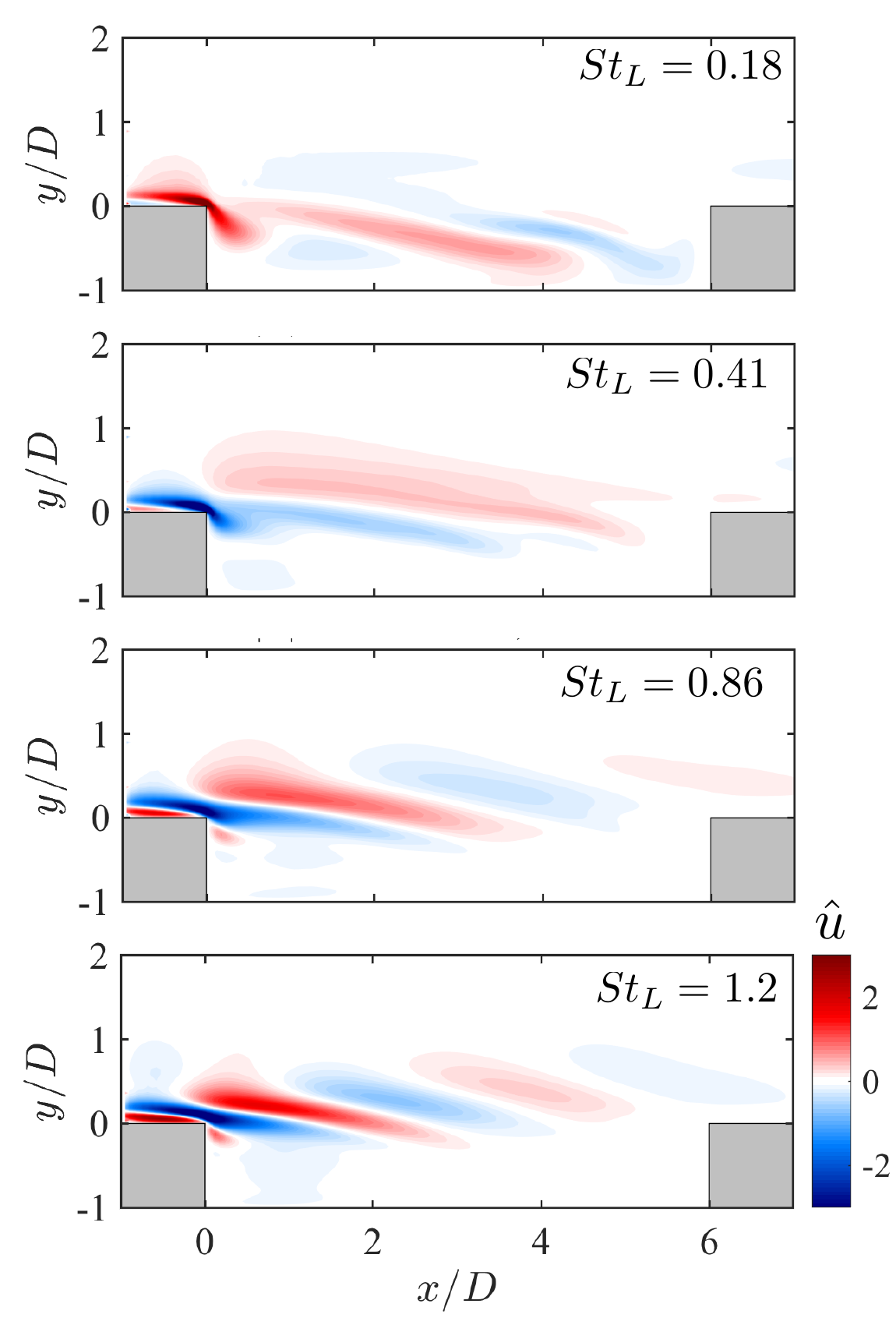}
	\caption{Streamwise velocity component $\hat u$ of forcing modes with $\beta=2$ and frequency over a range of $0.18\le St_L \le 1.2$ for the laminar cavity flow at $Re_D=502$.}
	\label{fig:forcing_low}
	\end{center}
\end{figure}

Similar influences of the frequency and wavenumber on forcing mode features are also observed. The forcing modes with $\beta=2$ and frequency in the range of $0.18\le St_L \le 1.2$ are shown in figure \ref{fig:forcing_low}. For all the frequencies, the largest value of disturbance in the forcing modes is around the cavity leading edge. When $St_L$ increases, the spatial structures of forcing modes also concentrate in the shear-layer region as observed for the response modes. Based on the spatial features of response and forcing modes, forcing with strong three dimensionality in the spanwise direction propagates in the shear layer region. We also observe that the flow inside the cavity does not respond to forcing with high frequencies ($St_L \gtrapprox 0.4$). 

For the turbulent flow at $Re_D=10^4$, there are unstable eigenmodes which were revealed by the stability analysis (figure \ref{fig:3D_eig}). For this reason, we perform the resolvent analysis with a temporal discount with $\alpha D/u_\infty=0.2$ satisfying $\alpha>\max({\tilde \omega_i})$. When a larger value of $\alpha$ is chosen, the finite-time window imposed on the system becomes narrower. As a consequence, the resolvent gain will be reduced, and the spatial extents that response and forcing modes cover become smaller in size since the modes are constrained to develop only within a short-time horizon.  On the other hand, this analysis highlights where the forcing and response modes originate in space. 

\begin{figure}[hbpt]
	\begin{center}
		\includegraphics[width=0.99\textwidth]{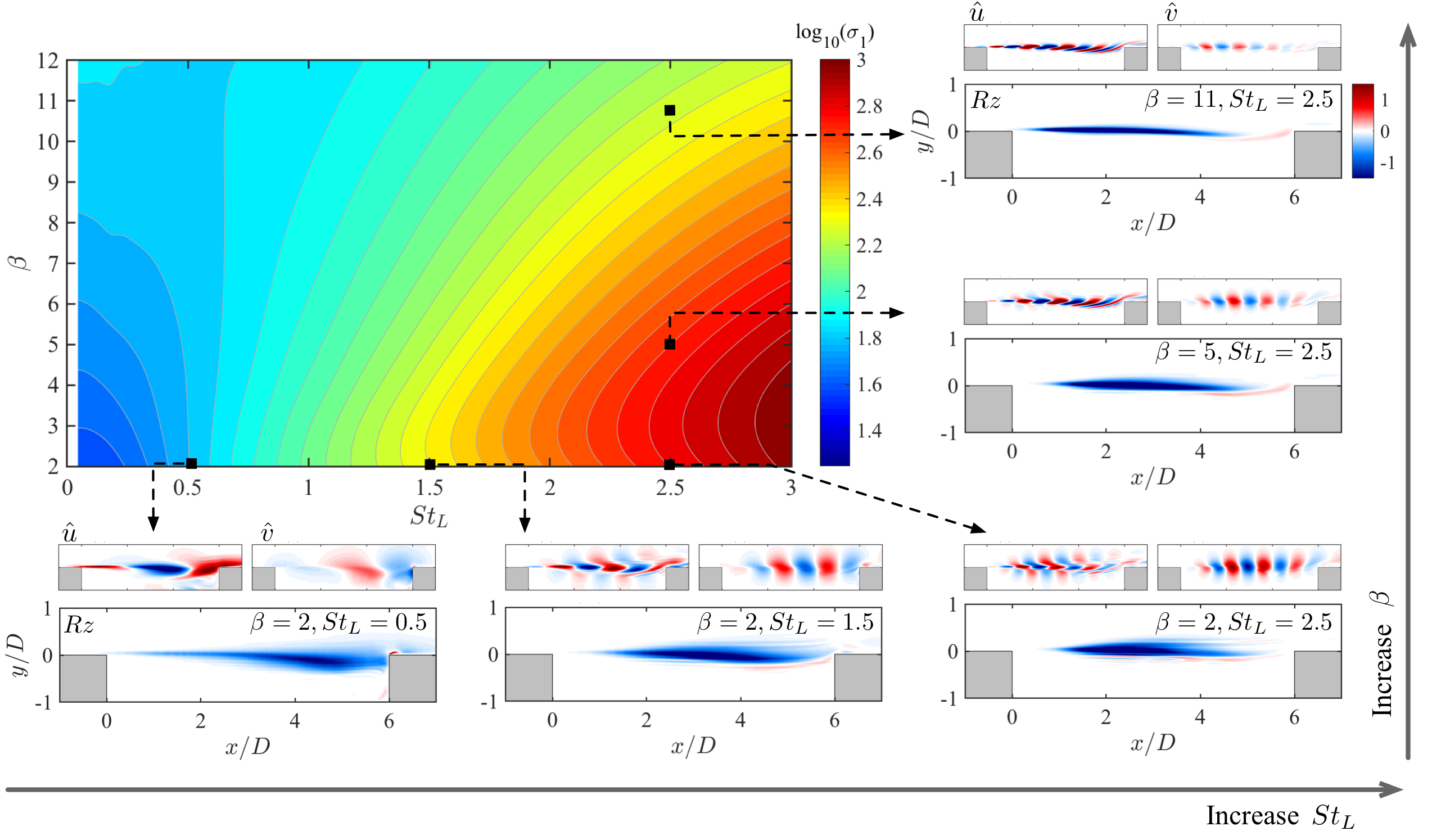}
	\caption{The leading gain map from resolvent analysis with $\alpha=0.12$ at $Re_D=10^4$. Subplots are representative response modes visualized using real component of streamwise velocity $\hat u$, transverse velocity $\hat v$ and Reynolds stress $R_z$.}
	\label{fig:3D_gain_high}
	\end{center}
\end{figure}

The frequency leading to high-resolvent gain is quite different between the turbulent and the laminar flows. The highest resolvent gain appears in high-frequency region with $St_L\gtrapprox2.5$ in the turbulent flow as shown in figure \ref{fig:3D_gain_high}, while the most responsive modes for the laminar cavity flow were associated with low frequency $St_L$. Moreover, the frequency of $St_L\gtrapprox2.5$ is much higher than the dominant Rossiter frequencies present in the baseline flows as presented in figure \ref{fig:2Dfreq}. The high gain phenomenon indicates the non-normality of the linear operator since there are no eigenmodes having the high frequencies (figure \ref{fig:3D_eig}) from stability analysis. For fluid-flow systems, large shear is generally responsible for a strong non-normality of the dynamical system. The structures of resolvent modes are not significantly affected whether the flow is laminar or turbulent, but the frequency at which the flow is most responsive differs. For the laminar cavity flow, lower frequency is preferred by the flow dynamical system, while higher frequency is more responsive in the turbulent flow. 

\begin{figure}[hbpt]
	\begin{center}
		\includegraphics[width=0.35\textwidth]{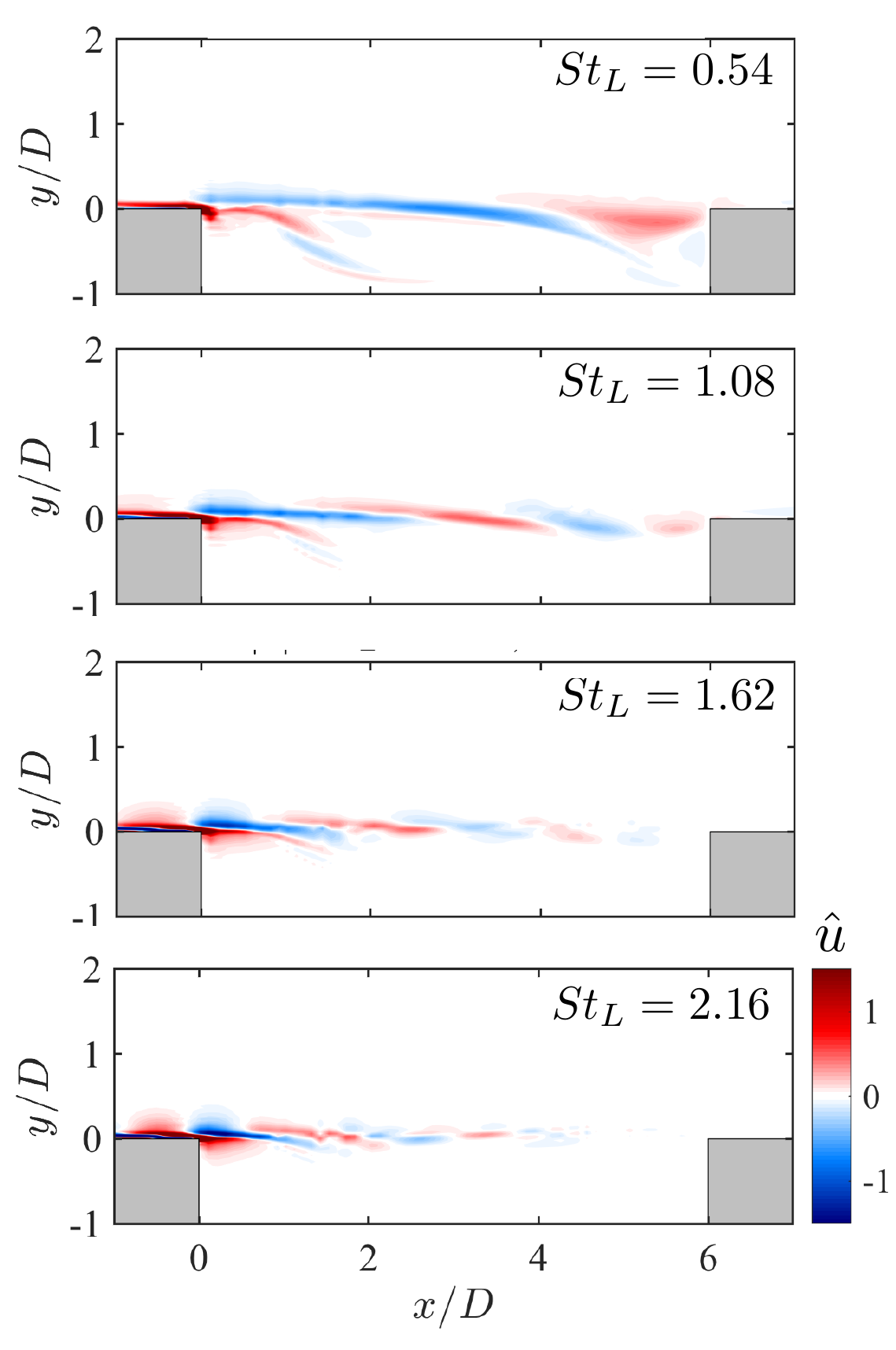}
	\caption{Streamwise velocity component $\hat u$ of the forcing modes with $\beta=9$ and frequency over a range of $0.54\le St_L \le 2.16$ for the turbulent cavity flow at $Re_D=10^4$.}
	\label{fig:forcing_high}
	\end{center}
\end{figure}

\begin{figure}[hbpt]
	\begin{center}
		\includegraphics[width=0.4\textwidth]{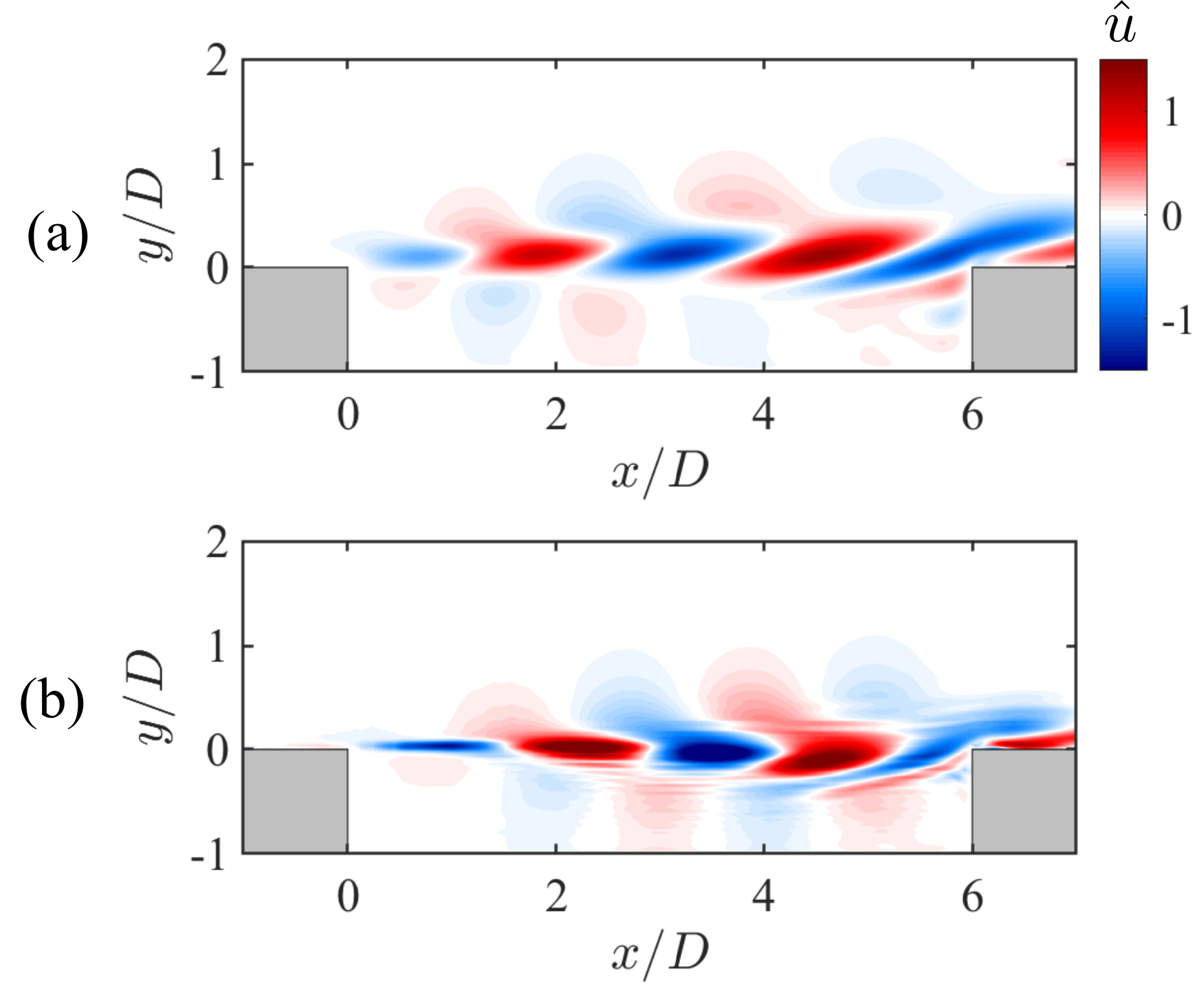}
	\caption{Comparison of the response modes with $\beta=2$ and $St_L=1.3$ for (a) laminar ($Re_D=502$) and (b) turbulent ($Re_D=10^4$) cavity flows. Contours show the real component of the streamwise velocity $\hat u$.}
	\label{fig:3D_mode_comp}
	\end{center}
\end{figure}

For turbulent flow, we find that the manner in which the response (figure \ref{fig:3D_gain_high}) and forcing modes (figure \ref{fig:forcing_high}) affected by frequency and spanwise wavenumber are similar to those of the laminar flow. For example, the structures of the 3D response modes with $St_L\gtrapprox0.5$ resemble the shear-layer instabilities. The spatial distribution of response mode tends to concentrate in the shear-layer region with a narrower vertical extent when $\beta$ increase. Larger Reynolds stress appears upstream in the shear layer for the higher frequency response modes. The largest value of disturbance in the forcing modes resides around the cavity leading edge.  A comparison of response modes from the laminar and the turbulent flows is shown in figure \ref{fig:3D_mode_comp}. The structures of the modes are comparable regardless of the huge difference in Reynolds number. It suggests that the coherent structures present in the flow are common structures across a wide range of Reynolds numbers. If a control strategy is designed based on the spatial information carried by the resolvent modes, its effective operating condition is likely to cover a wide range of Reynolds number for the same type of flow.  In terms of the parameter selection for control designs, those associated with high gains are good candidates with the least control effort required for the most amplified flow response. However, care must be taken if such mechanism leads to undesirable flow features. Those values of parameters should be avoided as studied in our companion work of controlling supersonic turbulent cavity flow guided by resolvent analysis.

\section{Summary}
\label{sec:sum}

In this paper, we applied resolvent analysis to study the input-output characteristics of open-cavity flows at a free stream Mach number of $M_\infty=0.6$. Both laminar and turbulent cavity flows at $Re_D=502$ and $10^4$, respectively, were considered. Since the main oscillation mechanism in the cavity flows is the shear-layer roll-up over a range of Reynolds numbers, the mean flow pattern is similar, especially in the shear-layer region, which is not significantly affected by the flow condition (laminar or turbulent). Two- and three-dimensional instabilities were captured by stability analysis, which provided a reference for choosing a proper finite-time window (discount parameter) for the resolvent analysis.      

With the resolvent analysis, we observed that 3D response modes with high frequency of $St_L\gtrapprox 0.5$ have similar spatial structures to the shear-layer instabilities. For a high frequency forcing in a 3D form around the cavity leading edge, the most responsive area in the flow is in the shear-layer region.  The spatial structures of the response modes share the same features for laminar and turbulent flows, and the scale of the response mode becomes smaller as frequency increases. Nevertheless, the larger the spanwise wavenumber is, the narrower the vertical extent of response mode. Although laminar and turbulent cavity flows have similar response modes in terms of the spatial structure, the overall energy norm amplification is much higher in the turbulent flow than in the laminar flow, and the preferred frequency revealed by high resolvent gain is different. The laminar flow is most responsive to low-frequency forcing with $St_L\lessapprox 0.4$, while the turbulent flow is most responsive to high-frequency forcing with $St_L\gtrapprox 2.5$. Therefore, the spatial structures of resolvent modes are insensitive to the flow state such that the features observed from fundamental laminar flow can be carried over to turbulent flow study. However, the amplification level in terms of energy norm and preferred frequency of the fluid dynamical system is strongly dependent on the flow state or Reynolds number.

\section*{Acknowledgments}

This work was supported by the Air Force Office of Scientific Research under award numbers FA9550-13-1-0091 and FA9550-17-1-0380 (Program Managers: Dr.~Douglas Smith and Dr.~Gregg Abate). YS, QL, and KT thank the insightful discussions with Dr.~Chi-An Yeh and the computational support from the Research Computing Center at the Florida State University.

\bibliographystyle{aiaa}
\bibliography{ref}


\end{document}